\documentclass[%
 aip,
 cha,
 amsmath,amssymb,
 reprint,%
a4paper
]{revtex4-2}

\usepackage{graphicx}%
\usepackage[utf8]{inputenc}
\usepackage[T1]{fontenc}
\usepackage{mathptmx}
\usepackage{hyperref}
\usepackage{microtype}
\usepackage[export]{adjustbox}

\begin{document}

\preprint{AIP/123-QED}

\title[]{The changing notion of chimera states, a critical review}

\author{Sindre W. Haugland}
 \email{sindre_w_haugland@mytum.de}
\affiliation{Physics Department, Nonequilibrium Chemical Physics, Technical University of Munich,
  James-Franck-Str. 1, D-85748 Garching, Germany}

\date{\today}%

\begin{abstract}
Chimera states, states of coexistence of synchronous and asynchronous motion, have been a subject of extensive research since they were first given a name in 2004.
Increased interest has lead to their discovery in ever new settings, both theoretical and experimental.
Less well-discussed is the fact that successive results have also broadened the notion %
of what actually constitutes a chimera state.
In this article, we critically examine how the results for different model types and coupling schemes, 
as well as varying implicit interpretations of terms such as coexistence, synchrony and incoherence, 
have influenced the common understanding of what constitutes a chimera. 
We cover both theoretical and experimental systems, address various chimera-derived terms that have emerged over the years and finally reflect on the question of chimera states in real-world contexts. 
\end{abstract}

\maketitle

\section{\label{sec:beginnings} Introduction}
\noindent Almost twenty years ago, Kuramoto and Battogtokh realized that a ring of coupled identical oscillators, when initialized appropriately, would remain %
divided into two distinguishable spatial regions~\cite{Kuramoto_NPCS_2002}: 
In one of these regions, the oscillators are %
mutually synchronized, while in the other, they drift with differing average frequencies.
Two years later, Abrams and Strogatz gave the new kind of symmetry-breaking its name, \emph{chimera state}, defining this as ``an array of identical oscillators split[ting] into two domains: one coherent and phase locked, the other incoherent and desynchronized''~\cite{Abrams_PRL_2004}.
Since then, oscillators with a common effective average frequency have been found to coexist with drifting oscillators in a variety of different models~\cite{Sethia_PRE_2013,Martens_PNAS_2013,Nkomo_PRL_2013,Omelchenko_Chaos_2015,Ulonska_Chaos_2016,Schmidt_PRE_2017,Maistrenko_Chaos_2020}.
However, the term ``chimera state'' has also come to be used to denote solutions wherein there is no such coexistence of different average frequencies. 
Among these are the time-periodic coupled-map-lattice solutions identified as one of the first experimental chimeras~\cite{Hagerstrom_NatPhys_2012}, along with the numerical chimeras %
inspiring them~\cite{Omelchenko_PRL_2011}. %
Similarly, in ``amplitude chimeras''~\cite{Zakharova_PRL_2014,Zakharova_JPhysCS_2016}, both ``coherent'' and ``incoherent'' oscillators orbit their respective average positions with the same average frequency.
Identified chimeras in ensembles of higher-dimensional (non-phase) oscillators sometimes defy a frequency-based classification altogether:
If the oscillators are aperiodic and 
more than one of the variables fluctuates sufficiently strongly, it becomes a non-trivial question 
how local frequencies might even be ascertained~\cite{Schmidt_Chaos_2014,Schmidt_Chaos_2015,Bohm_PRE_2015,Rohm_PRE_2016}. 

Underlying these examples is an implicit understanding of chimera states
less in terms of different long-term effective frequencies
than terms of the coexistence of one or more clusters (that is, individual units persistently possessing the mutually same value simultaneously) and several single, unclustered units.
This concept has also been invoked explicitly to predict the stability of chimera states in symmetric networks~\cite{Cho_PRL_2017}, as recently recognized in Ref.~\cite{OmelChenko_Nonlinearity_2018}.
A number of works~\cite{Schmidt_Chaos_2014,%
Kaneko_Chaos_2015,Kemeth_Chaos_2016,Hart_Chaos_2016} have further concluded that Kaneko observed chimera states in globally coupled logistic maps already towards the end of the 1980s~\cite{Kaneko_PhysicaD_1990}, 
presumably based on similar reasoning.
Because the ensemble consists of time-discrete
maps with chaotic dynamics, assigning them effective frequencies namely seems hardly possible.

\begin{figure*}[htb!]
    \centering
    \hspace{-4mm}
    \includegraphics[width=0.8\textwidth]{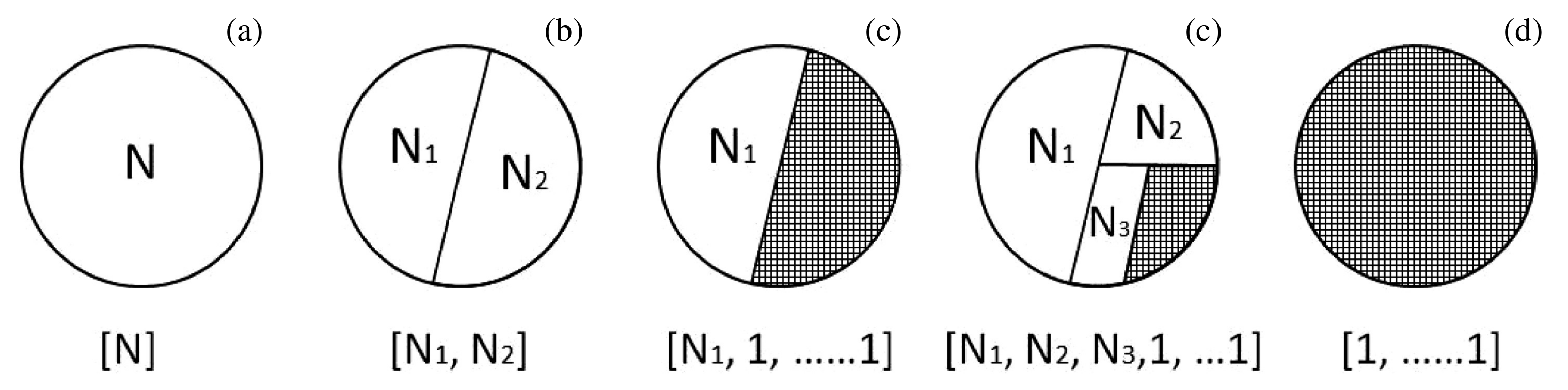}
    \caption{Schematic %
    examples of the different
    attractor types reported by Kaneko in 1989~\cite{Kaneko_PhysicaD_1990}. %
    Square brackets denote the sizes of clusters; letters (a-d) in the upper right denote the type of attractor according to the list in the introduction. 
    From left to right: $[N]$: Coherent (complete synchronization). $[N_1, N_2]$: Partition to two clusters as an example of attractor wherein the number of clusters does not grow with $N$. $[N_1, 1, 1,\dots, 1]$: Chimera state as an example of an attractor with $O(N)$ clusters, but at least one cluster comparable to $N$. $[N_1, N_2, N_3, 1, \dots, 1]$: A more complex attractor of the same fundamental type. %
    $[1, 1,\dots, 1]$: Complete desynchronization and no clusters that grow in size with $N$.
    Reprinted from K. Kaneko, Chaos 25 (9), 097608, 2015, with the permission of AIP Publishing. doi:\href{https://dx.doi.org/10.1063/1.4916925}{10.1063/1.4916925}, Ref.~\cite{Kaneko_Chaos_2015}.
    }
    \label{fig:Kaneko_Chaos_2015_chimera}
\end{figure*}

Indeed, it seems as if
Kaneko discovered both clustering and chimeras more or less simultaneously,
identifying a total of four different types of attractors, depending on
the sizes of the clusters they contain (including ``clusters'' of size 1).
These were 
\begin{enumerate}
\setlength{\itemsep}{0pt}
\item fully synchronized motion; as well as attractors with 
\item a~small number of clusters, much smaller than the system size $N$; 
\item a~large number of clusters in the order of $N$, but at least one cluster $N_1$ comparable in size to $N$; 
\item only small clusters in the order of 1.
\end{enumerate}
\noindent The still unnamed chimera state was only one among several different states of type (c). 
See Fig.~\ref{fig:Kaneko_Chaos_2015_chimera}.

Not long after, clustering was found and studied in globally coupled phase oscillators by Golomb et al.~\cite{Golomb_PRA_1992} and %
Okuda~\cite{Okuda_PhysicaD_1993}, and in globally coupled Stuart-Landau oscillators by Hakim and Rappel~\cite{Hakim_PRA_1992}.
Nakagawa and Kuramoto found a chimera state in the latter system in 1992, but did not treat it as more than a step from clustering to fully chaotic motion~\cite{Nakagawa_PTPS_1993}. 
Not before ten years later, after studying the nonlocally coupled CGLE and a ring of nonlocally coupled phase oscillators, did Kuramoto and Battogtokh publish the idea that the coexistence of synchronized and non-synchronized motion might be an interesting phenomenon in its own right~\cite{Kuramoto_NPCS_2002}.

Since then,
the number of reported chimeras has vastly increased, %
as summed up in several review papers%
~\cite{Panaggio_Nonlinearity_2015,Scholl_EPJST_2016,Bera_EPL_2017,OmelChenko_Nonlinearity_2018}.
Chimera states have also been found in ever more different models, meaning that the general conditions under which one or more of them could be said to occur, have become increasingly varied as well.
Over the course of this chimera research explosion, 
Schmidt and Krischer identified Kaneko's earliest chimera in 2013~\cite{Schmidt_Chaos_2014}, while
Sethia and Sen rediscovered Nakagawa and Kuramoto's 1992 chimera only few months later~\cite{Sethia_PRL_2014}.

There have also been made a few attempts to classify the increasing chimera-state variety:
In the appendix of their 2015 review, Panaggio and Abrams order the chimeras they have discussed according to system geometry and coupling type~\cite{Panaggio_Nonlinearity_2015}.
The appendix of Omel'chenko's 2018 review contains a similar ordering based on oscillator and coupling type~\cite{OmelChenko_Nonlinearity_2018}.
In 2016, Kemeth et al. published a possibly universal data-driven classification scheme, based on measures of spatial and temporal coherence, respectively~\cite{Kemeth_Chaos_2016}.
Yet, what no one seems to have pointed out so far, is how the entirety of states 
considered to be chimeras 
in the first place has 
evolved into
an ever more diverse collection of dynamical phenomena.
In addition to the aforementioned partial shift from a frequency to a clustering focus, 
this tacit liberalization of the term ``chimera state'' includes steps such as the extension from oscillators to maps, as well as ever new implicit interpretations of terms such as symmetric coupling, synchrony and incoherence. 
This review article critically examines the historical development of the notion of the chimera state, with an aim to let future research benefit from a more explicit understanding of how its object has changed.

Also noticeable is how Kaneko's original globally-coupled-maps chimera state was discovered and presented in the context of more or less related attractors right from the start. 
In contrast, 
Kuramoto and Battogtokh's chimera under nonlocal coupling was carefully constructed~\cite{Kuramoto_NPCS_2002} (for parameters where the fully synchronized solution is stable) and thus had little original context.
Context was instead created around it whenever new results in a wide range of models were produced, 
a process that has yielded terms such as ''alternating chimera``~\cite{Ma_EPL_2010}, ``multichimera''~\cite{Omelchenko_PRL_2013}, 
``chimera death''~\cite{Zakharova_PRL_2014} and ``weak chimera''~\cite{Ashwin_Chaos_2015},
to name but a few.
Seldom, though, is any particular system found to exhibit more than a few of these derived phenomena.
This issue 
will also be discussed below.

\section{From nonlocal to global coupling and various coupling schemes}
\noindent When the term ``chimera state'' was coined by Abrams and Strogatz in 2004, 
they cited only two prior works %
discerning this kind of previously unnamed dynamics~\cite{Abrams_PRL_2004}:
The first was a 2002 paper by Kuramoto and Battogtokh on its occurrence in the one-dimensional complex Ginzburg-Landau equation (CGLE) with nonlocal coupling.
Here, the coexistence of synchronized and unsynchronized oscillators was also found to persist as 
the originally complex-valued oscillatory medium was reduced to a phase-oscillator approximation~\cite{Kuramoto_NPCS_2002}. 
The second was a 2004 paper on the two-dimensional equivalent, a spiral wave with an incoherent core, found both in the phase approximation of the nonlocally coupled, spatially two-dimensional CGLE and in a 2D array of FitzHugh-Nagumo oscillators~\cite{Shima_PRE_2004}.

Kuramoto and Battogtokh considered their original chimera state to be ``among the variety of patterns which are characteristic to nonlocally coupled oscillators''~\cite{Kuramoto_NPCS_2002}.
In a 2006 paper, Abrams and Strogatz similarly draw the tentative conclusion that these dynamics are ``peculiar to the intermediate case of nonlocal coupling''~\cite{Abrams_IJBC_2006}.
The basis of their reasoning is two-fold:
Firstly, no other chimeras were known to them. 
Secondly, the coexistence of synchrony and incoherence in identical sine-coupled phase oscillators (the system to which Kuramoto and Battogtokh reduced their nonlocally coupled CGLE in 2002) does indeed become impossible when the
oscillators are coupled globally
~\cite{Abrams_IJBC_2006}. 
When Sethia and co-workers described chimera states in delay-coupled oscillators in 2008~\cite{Sethia_PRL_2008}, they were also under the impression that nonlocal coupling is indispensable.
The same goes for Wolfrum et al. during their 2011 investigations of the Lyapunov spectra and stability of chimera states of various sizes~\cite{Wolfrum_Chaos_2011,Wolfrum_PRE_2011}.

The mentioned papers all concentrated their attention on the case of weak coupling, where the CGLE can be approximated by phase oscillators, if not outright restricting themselves to phase oscillators as the starting point of their investigations.
So did a significant number of other papers published during the first decade of chimera-state research~\cite{Kawamura_PRE_2007,OmelChenko_PRL_2008,Abrams_PRL_2008,Ott_Chaos_2008,Pikovsky_PRL_2008,Laing_Chaos_2009,Laing_PhysicaD_2009,Sheeba_PRE_2009,Sheeba_PRE_2010,Bordyugov_PRE_2010,Martens_PRL_2010,Martens_PRE_2010,Martens_Chaos_2010,Shanahan_Chaos_2010,OmelChenko_PRE_2010,Ma_EPL_2010,Laing_Chaos_2012,Wildie_Chaos_2012}.
An exception was %
Laing's 2010 paper~\cite{Laing_PRE_2010}, which analyses a chimera state in an ensemble of Stuart-Landau oscillators without resorting to the phase reduction, but here too, the coupling is relatively weak and the amplitudes of individual oscillators deviate only a few percent from their average value.
Notably, all the non-synchronized oscillators are restricted to the same closed curve in the complex plane, which allows for effectively parametrizing their position by a phase alone, even though their amplitudes vary.

Only in 2013 did Sethia et al. show that a coexistence of coherent and incoherent dynamics can also occur in the nonlocally coupled CGLE in the case of strong coupling.
Here, the amplitude in the incoherent part of the system varies strongly in both space and time, inspiring the name ``amplitude-mediated chimeras'' (AMC) and preventing any attempt at a phase reduction~\cite{Sethia_PRE_2013}. 
Less than half a year later, Schmidt et al. successfully took the next step and identified a chimera state in an ensemble of Stuart-Landau oscillators with nonlinear \emph{purely global} (symmetric all-to-all) coupling~\cite{Schmidt_Chaos_2014}.
Like the AMCs of Sethia et al., this state exhibits strong amplitude fluctuations in the incoherent oscillators.
Moreover, it persists when adding diffusion and thus transitioning to a spatially extended 2D medium, where synchrony and incoherence form several clearly distinguishable intertwined islands. 
Shortly after, Sethia and Sen reported amplitude-mediated chimeras for Stuart-Landau oscillators with linear global coupling as well~\cite{Sethia_PRL_2014}.
In 2015, Laing complemented these efforts by producing chimera states in a ring of oscillators with only local (nearest-neighbor) coupling~\cite{Laing_PRE_2015}.

\begin{figure}[htb]
    \centering
    \includegraphics[width=0.8\columnwidth]{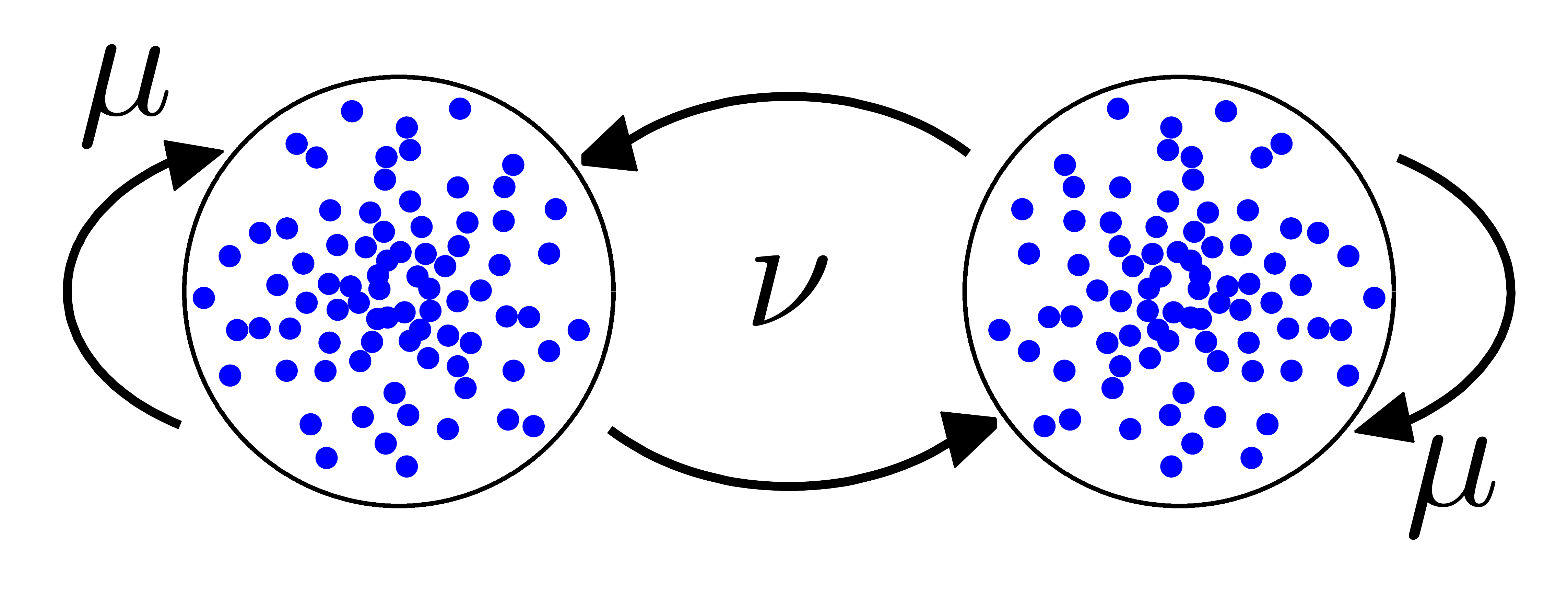}
    \vspace{-1mm}
    \caption{Schematic representation of the two-groups model introduced to the field of chimera states by Abrams and Strogatz in 2008~\cite{Abrams_PRL_2008} and inspiring a large number of subsequent works. 
    $\mu$ denotes the strength of the coupling within each group and $\nu$ that between the groups, usually with $\mu>\nu$.
     Reprinted figure with permission from M. Panaggio, D. M. Abrams, P. Ashwin \& C. R. Laing, Physical Review E 93 (1), 012218, 2016. doi:\href{https://dx.doi.org/10.1103/PhysRevE.93.012218}{10.1103/PhysRevE.93.012218}. Copyright 2016 by the American Physical Society.
    }
    \label{fig:Panaggio_PRE_2016_two_groups}
\end{figure}

The chimera states reported before the mentioned paper by Schmidt et al. had all been found in nonlocally coupled systems in the broad sense of the coupling being neither next-neighbor/diffusional (``local coupling'') nor all-to-all symmetric (``global coupling'').
However, the applied coupling schemes already encompassed a variety beyond Kuramoto and Battogtokh's original 1D ring topology and exponentially decaying coupling kernel.
Some of the alternative forms of nonlocal coupling were rather small variations, such as the use of a cosine kernel~\cite{Abrams_PRL_2004,Abrams_IJBC_2006} or a step function~\cite{OmelChenko_PRE_2010,Wolfrum_Chaos_2011} to limit the extent of the influence of each point in the system on the others.
(As long as the extent of the ring is restricted to $-\pi < x \leq \pi$, the cosine kernel $G(x - x') \propto 1 + A \cos(x-x')$, and thus the coupling strength, also decreases monotonously with distance~\cite{Abrams_PRL_2004}.)
More radical was the idea of dividing the identical oscillators into two populations with symmetric all-to-all coupling within each group,
as well as a weaker coupling between the 
groups~\cite{Abrams_PRL_2008}.
See Fig.~\ref{fig:Panaggio_PRE_2016_two_groups}.
Besides Kuramoto and Battogtokh's original system, this ``simplest network of networks''~\cite{Laing_PRE_2010}, is possibly the most influential theoretical model supporting chimera states, and similar models have been the subject of a large number of subsequent works~\mbox{\cite{Pikovsky_PRL_2008,Laing_Chaos_2009,Sheeba_PRE_2009,Ma_EPL_2010,Martens_PRE_2010,Martens_Chaos_2010,Laing_PRE_2010,Sheeba_PRE_2010,Laing_Chaos_2012,Laing_Chaos_2012a,Tinsley_NatPhys_2012,Wildie_Chaos_2012,Martens_PNAS_2013,Pazo_PRX_2014,Buscarino_PRE_2015,Panaggio_PRE_2016,Martens_Chaos_2016,Martens_NJP_2016}}%

While the 2008 paper by Abrams et al. does not cite any prior source for the two-groups model, Laing points out in Ref.~\cite{Laing_Chaos_2009} that more general versions of it were actually implemented in earlier articles by Montbrió et al.~\cite{Montbrio_PRE_2004} and Barreto et al.~\cite{Barreto_PRE_2008}.
Here, the system studied in Montbrió's 2004 article differs only from the later system of Abrams et al. by the fact that the oscillators (within each group) are given heterogeneous natural frequencies.
The intra-group coupling is the same within each group, and the inter-group coupling is the same in either direction.
Under these circumstances, the authors most notably find generalized versions of what Abrams et al.~\cite{Abrams_PRL_2008} would later call ``stable chimera'' and ``breathing chimeras''.
(See section~\ref{sec:chimera_types}.)
The first to draw attention to this seem to have been Laing in two 2009 articles, wherein he, in addition to expanding the chimera notion to heterogeneous oscillators, also expanded the mathematical foundations of these generalized chimeras.

In a different variation from the simplest two-groups model, Laing gradually removes connections between oscillators in order to further test the robustness of the chimera, and finds it to be more sensitive to removal of intra-group than of inter-group links~\cite{Laing_Chaos_2012}.
Like Laing's use of 
heterogeneous frequencies, this removal of connections
is motivated by the aim to create a more realistic model, 
as neither really identical oscillatory units nor perfectly symmetric coupling schemes are likely to exist in nature~\cite{Laing_Chaos_2009,Laing_PhysicaD_2009,Laing_Chaos_2012}.

Among the other chimera-supporting systems inspired by the two-groups model were networks of three~\cite{Martens_PRE_2010,Martens_Chaos_2010} and eight populations~\cite{Wildie_Chaos_2012}, respectively.
Additional variants 
include one in which the individual links between the populations are randomly switched on and off at equally spaced time intervals~\cite{Buscarino_PRE_2015}, 
one with different phase lags in the coupling within and between populations~\cite{Martens_Chaos_2016}, 
and a delay-coupled version with different intra- and inter-population coupling delays~\cite{Sheeba_PRE_2009,Sheeba_PRE_2010}.
Schmidt et al. also invoke the two-groups model when explaining the stability of their globally coupled chimera.
Here, the synchronized and incoherent oscillators effectively form two self-organized groups, and these groups exert different influences on the respective other group, thereby further reinforcing the chimera state~once~spontaneously~formed~\cite{Schmidt_Chaos_2014}.

\section{Types of chimera states and chimera-derived concepts}
\label{sec:chimera_types}
\noindent The gradual expansion of the general concept of a chimera state was accompanied by the naming of an increasing number of derived phenomena, among them the aforementioned amplitude-mediated 
and amplitude chimera 
states.
Discerned already in 2008 was the ``breathing chimera''.
Here, the phase coherence of the incoherent oscillators, quantified by the order parameter $r(t) = |\langle \mathrm{e}^{i \theta_j(t)} \rangle_\mathrm{incoh.}|$, where the sum is taken over the phases $\theta_j$ of all oscillators in the unsynchronized group, is either periodic~\cite{Abrams_PRL_2008} or quasiperiodic~\cite{Pikovsky_PRL_2008}. 
This contrasts with what Abrams et al. call a ``stable chimera''~\cite{Abrams_PRL_2008}, such as the one discovered by Kuramoto and Battogtokh, where $r(t)$ is constant in time.
A few later works have also used the term ``breathing chimera'' to denote a chimera in which the coherent and incoherent parts move through the system, while the global degree of clustering might remain constant throughout~\cite{Sethia_PRE_2013,Vuellings_NJP_2014}, possibly because this makes the \emph{local} order parameter ``breathe''.
In 2014, the periodic and quasiperiodic chimeras were complemented by a chimera state wherein the order parameter behaves chaotically, identified by Pazo et al. in a two-groups system of identical Winfree oscillators~\cite{Pazo_PRX_2014}.

\begin{figure*}[ht!]
\hspace{-10mm}
    \includegraphics[width=1.6\columnwidth]{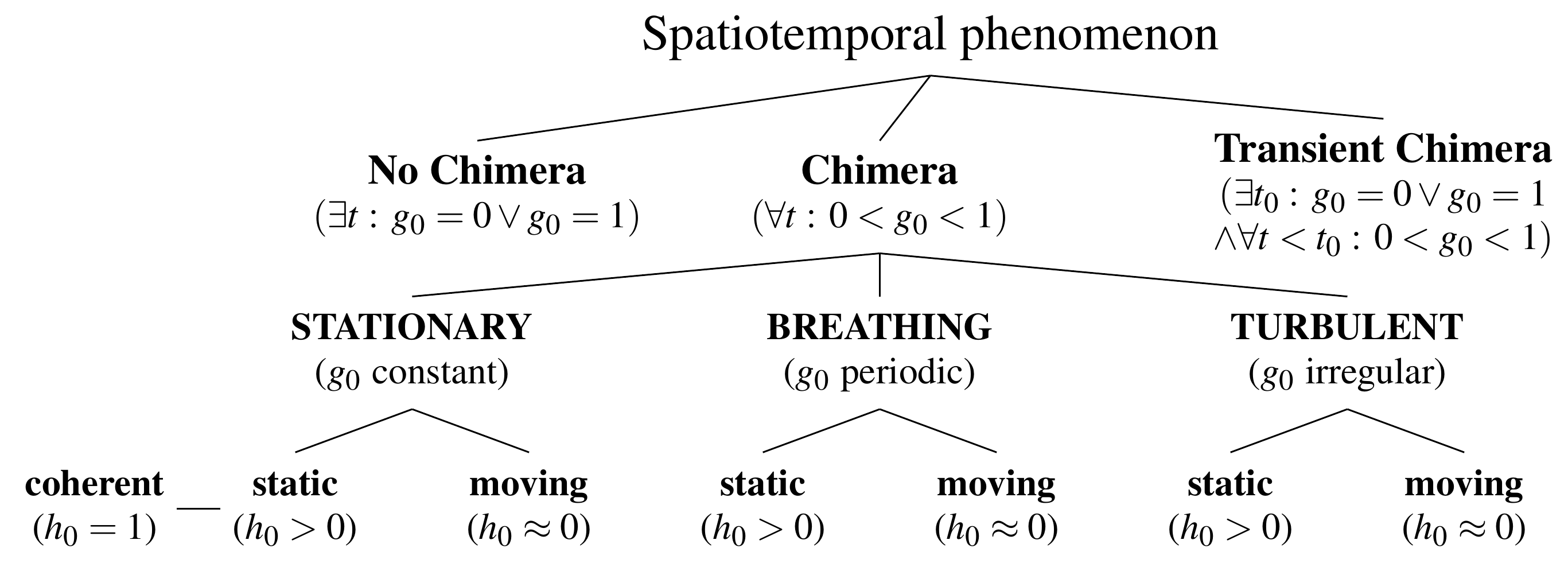}
    \caption{Data-driven classification scheme of %
    Kemeth et al.%
    ~\cite{Kemeth_Chaos_2016}.
    Any spatiotemporal phenomenon is assessed in the form of time-series vectors representing the different parts or units (e.g. oscillators, maps) of the system.
    For each time $t$, the measure $g_0 \in [0, 1]$ indicates how similar on average the value of each unit is to either its neighbors (in spatially ordered systems) or all other units.
    The measure $h_0 \in [0, 1]$ indicates the fraction of the units that are strongly temporally correlated over the evaluated interval.
    Phenomenologically, the synchronized part of the system remains fixed in ``static chimeras'', while it moves in ``moving chimeras.''
    }
    \label{fig:Kemeth_Chaos_2016_classification_scheme}
\end{figure*}

Also coined in 2008 was the term ``clustered chimera state'' with several coherent regions phase-shifted relative to each other~\cite{Sethia_PRL_2008}. 
This was joined five years later by the concept of the ``multichimera'', likewise containing several distinct coherent regions, but with no phase difference between them~\cite{Omelchenko_PRL_2013}.
Either of these two phenomena have since come to be called both multicomponent~\cite{Ujjwal_PRE_2013}, multicluster~\cite{Xie_PRE_2014,Yao_SciRep_2015}, multiheaded~\cite{Maistrenko_IJBC_2014,Larger_NatCom_2015} or multiple-headed chimera~\cite{Larger_NatCom_2015,Schmidt_PRE_2017}.
In the case of equal coupling to a fixed number of nearest neighbors (that is, a step-function nonlocal coupling), the number of incoherent regions (referred to as ``heads''~\cite{Maistrenko_IJBC_2014}) increases if the coupling \emph{range} is made shorter~\cite{Omelchenko_PRL_2013,Vuellings_NJP_2014,Omelchenko_Chaos_2015}.
If the coupling \emph{strength} is made stronger, the number of heads may either increase~\cite{Omelchenko_PRL_2013} or decrease~\cite{Omelchenko_Chaos_2015}, depending on the underlying type of oscillator.
More complex coupling topologies can also produce various multiplicities of coherent and incoherent regions, but the determining factors are less obvious there~\cite{Hizanidis_PRE_2015,Ulonska_Chaos_2016}.
A 2D equivalent of the 1D multichimera is the multicore spiral chimera reported by Xie et al.~\cite{Xie_PRE_2015}.
In the chimera found by Schmidt et al. in the CGLE with nonlinear global coupling in 2013~\cite{Schmidt_Chaos_2014}, the concept of a distinct number of chimera heads is less meaningful, as synchronized and incoherent regions, respectively, merge with time~\cite{Haugland_SciRep_2015}. 

In 2015, Ashwin and Burylko proposed a rigorous chimera definition applicable to small ensembles.
They did this by defining a ``weak chimera'' to be a state in which the average phase velocities of at least two oscillators converge in the limit of infinite time $T\rightarrow \infty$, while it remains different for at least one other oscillator~\cite{Ashwin_Chaos_2015}.
This definition was subsequently used to classify states in several later works~\cite{Suda_PRE_2015,Omelchenko_PRL_2016,Panaggio_PRE_2016,GarciaMorales_EPL_2016,Maistrenko_PRE_2017,Kemeth_PRL_2018,OcampoEspindola_FAMS_2019}.
One may assume that it was inspired by the observation that the effective average frequencies of the incoherent oscillators differ from the average frequency of the synchronized cluster in both Kuramoto and Battogtokh's 2002 chimera~\cite{Kuramoto_NPCS_2002} as well as several later chimera states~\cite{Omelchenko_PRL_2013,Nkomo_PRL_2013,Sethia_PRE_2013,Maistrenko_IJBC_2014,Vuellings_NJP_2014,Rosin_PRE_2014}.
However, not all %
identified classical chimera states (in the sense of some kind of coexistence of synchrony and incoherence) are actually weak chimeras as well. 
In particular they cannot be when the incoherent region drifts through the system with time and, as a consequence, all oscillators take turns being either coherent or incoherent~\cite{Ashwin_Chaos_2015}. 
Not long after Ashwin and Burylko published their definition, Panaggio et al., for the purposes of their paper on two small oscillator populations, used 
it to define a chimera (without any qualifications) to mean a weak chimera in which the frequency-synchronized oscillators have the same phase~\cite{Panaggio_PRE_2016}.
Two years later, findings by Kemeth et al. implicitly challenged the potential use of this as a general definition: 
when scaled up, the two unsynchronized oscillators of a certain %
minimal (weak) chimera with a perfectly synchronized coherent part are namely not replaced by a greater number of incoherent oscillators~\cite{Kemeth_PRL_2018}.
Instead, the state becomes a three-cluster solution with one large and two small clusters.
This contrasts with a different kind of minimal chimera that the authors also identify, wherein the dimensionality of the dynamics grows with the system size and which they thus coin an ``extensive chimera state''~\cite{Kemeth_PRL_2018}.

Also notable is the ``alternating chimera'', in which two equivalent parts of a system 
take turns being synchronized and incoherent, respectively.
This was first produced by external periodic forcing~\cite{Ma_EPL_2010} and later found to arise autonomously, in two pre-defined populations~\cite{Laing_Chaos_2012} as well as in a globally coupled oscillatory medium
~\cite{Haugland_SciRep_2015}. 
Other chimera-inspired terms include the ``globally clustered chimera'', denoting a system of several pre-defined populations that all split into both synchronized and incoherent oscillators~\cite{Sheeba_PRE_2009};
``chimera death'', the coexistence of spatially coherent and incoherent oscillation death~\cite{Zakharova_PRL_2014,Zakharova_JPhysCS_2016};
and the poetically named ``Bellerophon states'' that occur when a certain chimera state is made unstable by parameter tuning~\cite{Qiu_SciRep_2016}.
The ``solitary state'' is a chimera that does not contain a spatially continuous incoherent region, but wherein one or more single oscillators are desynchronized from an otherwise homogeneous background~\cite{Maistrenko_PRE_2014,Jaros_PRE_2015,Maistrenko_Chaos_2020}.

Additional chimeric phenomena are 
the ``turbulent chimera''~\cite{Bordyugov_PRE_2010}, the ``intermittent chaotic chimera''~\cite{Olmi_PRE_2015}, the ``scroll ring chimera'' (in three spatial dimensions)~\cite{Maistrenko_NJP_2015,Maistrenko_Chaos_2020} 
and the ``blinking chimera''~\cite{Goldschmidt_Chaos_2019},
as well as the ``antichimera'' and ``dual chimera''~\cite{Petrungaro_PRE_2017}.
In contrast, Laing in a 2012 paper~\cite{Laing_Chaos_2012} reported a kind of imperfect chimera in which one half of the oscillators are more strongly clustered than the other, while none of the two groups is fully synchronized, but without giving this phenomenon an additional name.

When Kemeth et al. came up with a general classification scheme for chimera states in 2016, the wide variety of both existing chimeras and the systems in which they occur made the authors pick a data-driven approach~\cite{Kemeth_Chaos_2016}. 
Their scheme, reproduced in Fig.~\ref{fig:Kemeth_Chaos_2016_classification_scheme} uses some of the aforementioned labels, such as breathing~\cite{Abrams_PRL_2008,Pikovsky_PRL_2008,Sethia_PRE_2013} and turbulent chimera~\cite{Bordyugov_PRE_2010}, in addition to coining new terms, such as ``moving chimera'' and ``static chimera''.
In a moving chimera, most individual constituent units of the regarded system change from incoherent to synchronized or vice versa within the regarded time interval, 
while in a static chimera, they do not~\cite{Kemeth_Chaos_2016}.
Notably, 
the authors did not only apply their classification scheme to already declared chimera states, but to other dynamics as well. 
These include the 
localized turbulence in the CGLE with time-delayed linear global coupling, as reported by Battogtokh et al. already in 1997~\cite{Battogtokh_PhysicaD_1997}.
It is classified as a ``turbulent moving chimera'',
which differs somewhat from the earlier conclusion of Schmidt et al., who, on finding localized turbulence in the CGLE with nonlinear global coupling, were more reluctant to call it a chimera state~\cite{Schmidt_Chaos_2015}.
Kemeth et al. also evaluate the gradual formation of incoherent patches on a uniformly oscillating background in Falcke and Engel's 1994 work on a model of CO coverage on a platinum surface~\cite{Falcke_PRE_1994,Falcke_JChemPhys_1994,Falcke_Thesis_1995}.
The data-based classification scheme groups these dynamics in the same category of finite-lifetime states as the aforementioned amplitude chimera, a category Kemeth et al. suggest to call ``transient chimera''.
While not covered by the article on the classification scheme, what Yang et al. call ``localized irregular clusters'' in a 2000 paper on the Belousov-Zhabotinsky reaction with global feedback~\cite{Yang_PRE_2000} also looks suspiciously like a chimera state.

\section{Experimental chimeras and a broader chimera concept}
\noindent One of the very first experimental realizations of a chimera state was also based on the two-groups model.
It was a 2012 photochemical experiment by %
Tinsley et al.,
involving 40 photosensitive Belousov-Zhabotinsky (BZ) oscillator beads.
Here, each of the beads emits light of a certain intensity, which is recorded with a CCD camera and projected selectively back on the beads by a spatial light modulator (SLM)~\cite{Tinsley_NatPhys_2012,Note1}%
.
About a year later, the two-groups model also formed the basis for a purely mechanical chimera, without any computer-mediated coupling, found in metronomes placed on two swings connected by springs: 
The intra-group coupling is conveyed by vibrations of the respective swing, while inter-group coupling happens via the springs~\cite{Martens_PNAS_2013}.
Published by Hagerstrom et al. simultaneously with the BZ chimera paper, 
but inspired by a different model~\cite{Omelchenko_PRL_2011}, 
was an optical realization of an array of coupled maps~\cite{Hagerstrom_NatPhys_2012}: 
Here, the different parts of the cross-section of a beam of circularly polarized light have different phases when they emerge from an SLM. 
As the beam passes through an optical setup, these phases are translated into intensities recorded by a camera, and these intensities in turn determine which phase shift the SLM is to apply to each part of the beam in the next iteration.

While the first laboratory chimeras had taken a full ten years since Kuramoto and Battogtokh's 2002 chimera state~\cite{Smart_PhysToday_2012}, the next few years saw a much more rapid addition of experimental realizations, including mechanical models~\cite{Martens_PNAS_2013,Olmi_PRE_2015,Wojewoda_SciRep_2016}, networks of discrete electrochemical oscillators~\cite{Wickramasinghe_PlosOne_2013} 
and various electronic and optoelectronic systems~\cite{Gambuzza_PRE_2014,Bohm_PRE_2015,Larger_NatCom_2015,Hart_Chaos_2016,Brunner_Chaos_2018}.
The first experimental chimera in a system with global coupling seems to have been observed in 2013 in an photoelectrochemical setup
~\cite{Schmidt_Chaos_2014,Schoenleber_NJP_2014}.
Also of particular interest is the experimental chimera published by Totz et al. in 2017~\cite{Totz_NatPhys_2017}:
Here, BZ beads of the type previously used by Tinsley et al.~\cite{Tinsley_NatPhys_2012} are coupled by means of the same kind of optical feedback and virtually arranged in a $40\times40$ grid.
For suitable experimental parameters, this yields a spiral-wave chimera with an incoherent core -- the qualitatively same kind of pattern as the first 2D numerical chimeras published by Shima and Kuramoto in 2004~\cite{Shima_PRE_2004}. 

Several of the experimental chimeras published from 2012 onward differ significantly from what a chimera state had originally been:
Already one of the first experimental realizations had worked with chaotic maps~\cite{Hagerstrom_NatPhys_2012}, and not with oscillators, as it said in Abrams and Strogatz' 2004 definition, but this might just be taken as an extension of the phenomenon to a new domain.
More remarkable in light of the original definition
is the fact that this coupled-maps chimera 
is (partially) incoherent only in space, while the whole system is temporally periodic~\cite{Hagerstrom_NatPhys_2012}.
Something similar applies to the so-called ``chimera states with quiescent and synchronous domains'' found in the coupled electronic oscillators of Gambuzza et al. two years later~\cite{Gambuzza_PRE_2014}:
Here, the voltage of some constituent circuits is constant in time, while it is oscillating with the same frequency in all the others, but there is no desynchronized region.
Arguably also softening the original concept was the mechanical chimera state reported by Wojewoda et al. in 2016. 
Here,
two out of only three coupled 
pendula are synchronized, while the third one, which ends up oscillating
uncorrelated with the oscillation of the other two, is declared to constitute its own incoherent group~\cite{Wojewoda_SciRep_2016}.
Similarly unprecedented was the optoelectronic chimera published by Larger et al. one year before, in which there are no physically distinct coupled units, but just a single semiconductor laser setup subject to time-delayed feedback~\cite{Larger_NatCom_2015}: 
The result is a single time-varying signal, with features (among others) on the length-scale of the applied delay; only when this signal is chopped up into segments and these segments are stacked on top of each other to form the temporal evolution of a ``virtual space'' does the chimera state appear to the observer.
In 2018, Brunner et al. successfully repeated the same procedure for a setup with two simultaneously applied delays of different magnitude, thereby creating a chimera in 2D virtual space~\cite{Brunner_Chaos_2018}.

Actually, this broadening of the scientific community's chimera concept had already begun in the theoretical systems:
Already in 2008 did Omel'chenko et al. investigate a 1D array
in which the force on each particular oscillator is not only proportional to its deviation from the common mean, but also dependent on its absolute location in the array~\cite{OmelChenko_PRL_2008}.
Such a coupling scheme is \emph{not} symmetric in the sense that all oscillators are governed by the same equation of motion and would all feel the same force if they were fully synchronized.
The resultant coexistence of synchrony (where the spatial modulation is strong) and incoherence (where the spatial modulation is weak) was nevertheless declared to be a chimera state.
Something similar applies to Laing's later gradual and random removal of individual links from the two-groups model~\cite{Laing_Chaos_2012} and to the randomly time-varying links which Buscarino et al. published in~2015~\cite{Buscarino_PRE_2015}.

Another chimera state, recognized in coupled maps 
by Omelchenko et al. in 2011, is notably 
periodic in time and incoherent only in space~\cite{Omelchenko_PRL_2011}
(thereby preceding the experimental chimera of Hagerstrom et al.~\cite{Hagerstrom_NatPhys_2012} in this regard).
This breaks with the part of Abrams and Strogatz' original chimera definition~\cite{Abrams_PRL_2004} that assigns the attribute of being phase locked to the coherent group only.
By allowing for chimera states in coupled maps, Omelchenko et al. 
also laid the foundations for the later recognition of Kaneko's much earlier globally-coupled-maps chimera~\cite{Kaneko_PhysicaD_1990}.

\begin{figure}[htb]
    \centering
    \includegraphics[width=0.95\columnwidth]{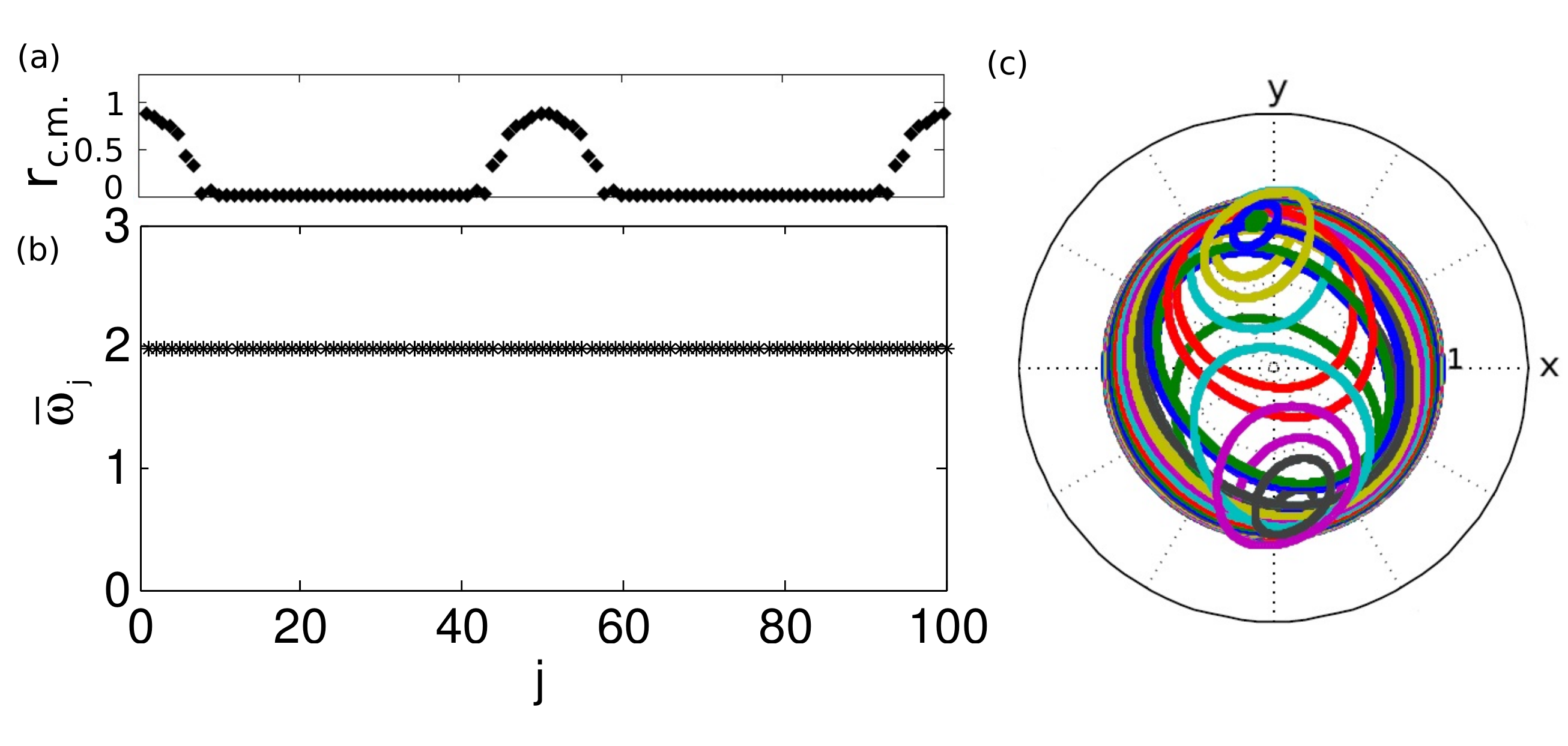}
    \vspace{-2mm}
    \caption{Amplitude chimera in a ring of nonlocally coupled Stuart-Landau oscillators~\cite{Zakharova_JPhysCS_2016}.
    (a) Averaged over a period, the position of all oscillators along the two ``coherent'' segments of the ring 
    is centered on the origin, 
    while that  
    along the two ``incoherent'' ring segments form an arc-like shape, reminiscent of the distribution of average frequencies in the incoherent part of Kuramoto and Battogtokh's 2002 chimera.
    (b) %
    The average phase velocity $\overline{\omega}_j$ with which each oscillator orbits its own respective average position, is the same throughout. 
    (c) Phase portrait of all oscillators in the complex plane.
    Reprinted from A. Zakharova, M. Kapeller \& E. Schöll, Journal of Physics: Conference Series 727 (1), 012018, 2016. doi:\href{https://dx.doi.org/10.1088/1742-6596/727/1/012018}{10.1088/1742-6596/727/1/012018} under the terms of the Creative Commons Attribution 3.0 licence.
    }
    \label{fig:Zakharova_JPhysCS_2016_amplitude_chimera}
\end{figure}

Similarly expanding the definition of chimera states were the ``amplitude chimeras'' of Zakharova et al.~\cite{Zakharova_PRL_2014,Zakharova_JPhysCS_2016}: 
Here, all oscillators oscillate in synchrony, with the ``incoherent'' ones oscillating around different points in the complex plane than the ``coherent'' ones, 
as well as having different radii of oscillation.
See Fig.~\ref{fig:Zakharova_JPhysCS_2016_amplitude_chimera}.
Some of the cellular-automaton chimeras published by 
Garc\'{i}a-Morales in 2016 also have a well-known periodicity; 
and while the author found no periodicity for some of the others during the tested simulation time, it is of course fundamentally true that ``because of the finiteness of the dynamics, the periodicity of any structure is bounded''~\cite{GarciaMorales_EPL_2016},
that is, because the states of the system are discrete, it is bound to repeat itself eventually.
Garc\'{i}a-Morales was possibly also the first to %
recognize how the community's chimera definition had broadened, %
declaring to ``regard chimera states as an experimental fact of nature rather than a feature of certain systems of differential equations or maps''~\cite{GarciaMorales_EPL_2016}. %
At about the same time, Bastidas et al.~\cite{Bastidas_PRE_2015,Bastidas_Book_2016} proposed to extend the chimera notion to quantum mechanics through work on coupled quantum van der Pol oscillators.

\begin{figure*}[htb]
    \centering
    \includegraphics[width=1.5\columnwidth]{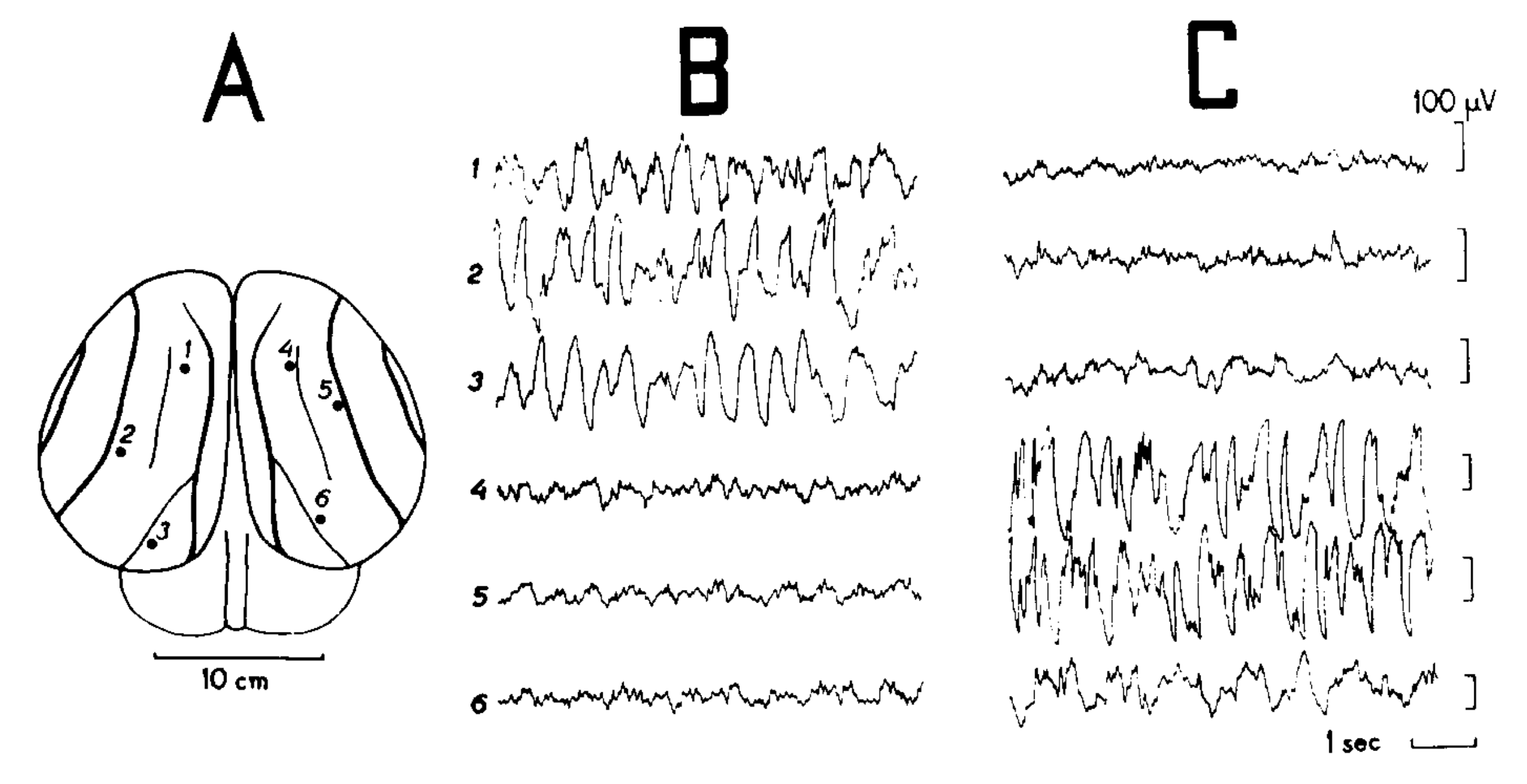}
    \vspace{-1mm}
    \caption{
    EEG measurements of the bottlenose dolphin %
    published by Mukhametov et al. in 1977~\cite{Mukhametov_BrainRes_1977} and later reprinted by Rattenborg et al.~\cite{Rattenborg_NeuBioRev_2000}.
    (a)~Location of the electrodes across the two brain hemispheres.
    (b,c) EEG activity measured by each electrode during two short intervals recorded one hour apart.
    The sleeping hemisphere (electrodes 1-3 in b and 4-6 in c) is characterized by high-amplitude low-frequency EEG activity, the awake hemisphere by low-amplitude high-frequency EEG activity.
     Reprinted figure with permission from L. M. Mukhametov, A. Y. Supin \& I. G. Polyakova, Brain Research 134 (3), 581-584, 2016. doi:\href{https://dx.doi.org/10.1016/0006-8993(77)90835-6}{10.1016/0006-8993(77)90835-6}. Copyright 1977 Published~by~Elsevier~B.V.
    }
    \label{fig:Mukhametov_BrainRes_1977_unihemispheric}
\end{figure*}

\section{Different standards for natural-world chimeras?}
\noindent Any broader treatment of chimera states
should reflect at least briefly on the possibility of chimeras outside of laboratories.
Here, the phenomenon most readily invoked by the 
community is probably unihemispheric sleep, with Rattenborg, Amlaner and Lima's extensive 2000 neuroscientific review paper~\cite{Rattenborg_NeuBioRev_2000} being cited in the introductions of many of the studies mentioned above.
Very briefly explained, a unihemispherically sleeping animal sleeps with one half of its brain at a time.
Aquatic mammals sleep this way, allowing them to surface to breathe, as do birds and at least some reptiles~\cite{Mathews_Ethology_2006}.
When measured, the EEG activity in the sleeping brain hemisphere is high-amplitude, low-frequency, while that in the awake hemisphere is low-amplitude, high-frequency, implying that the individual neurons in the former are firing more strongly synchronized than those in the latter~\cite{Rattenborg_NeuBioRev_2000}.
First to notice the possible connection to early numerical chimeras were probably Abrams et al., whose 2008 chimera-supporting two-groups model is motivated by the question of what might be the simplest system of two oscillator populations to emulate this kind of brain behavior~\cite{Abrams_PRL_2008}.

While the roles of the synchronized and incoherent group in the two-groups model are fixed once established, natural unihemispheric sleep tends to move from one half of the brain to the other several times over the course of an interval of sleeping. 
This was recognized by Ma, Wang and Liu in a 2010 paper, wherein they describe 
the first of the aforementioned alternating chimera states. 
However, in order to observe the switching of synchrony from one population to the other, they have to resort to an external periodic forcing, which they declare to ``represent the varying environment''~\cite{Ma_EPL_2010}.
In 2015, Haugland et al. reported a fully self-organized alternating chimera without neither pre-defined groups nor any external force, claiming to ``tighten~[\dots] the connection between chimera states and unihemispheric sleep''~\cite{Haugland_SciRep_2015}.
As recently as 2019, chimera states have also been realized in two different two-layer networks more closely inspired by brain architecture~\cite{Ramlow_EPL_2019,Kang_SciRep_2019}, thereby
completing the continuum of phenomena from 
actual unihemispheric sleep to the most ideal mathematical chimera.

Related to unihemispheric sleep and also mentioned as a motivation for chimera research is the ``first-night effect'' in humans, keeping one hemisphere more vigilant when sleeping in a novel environment~\cite{Tamaki_CurrBio_2016,Ramlow_EPL_2019}.
Other authors have likened chimera states to the regional highly synchronized brain activity during epileptic seizures or due to Parkinson's disease~\cite{Sethia_PRL_2014}.
Similarly, spiral-wave chimeras, with their incoherently fluctuating core~\cite{Shima_PRE_2004}, have been compared to ventricular fibrillation, when rotating patterns of excitation occur on the heart, with possibly uncoordinated dynamics at their center~\cite{Panaggio_Nonlinearity_2015}.
But neither of these comparisons seem to have sparked in-depth deliberation like unihemispheric sleep.

In their 2013 article on chimera states in two pendulum populations, Martens et al. claim that their ``model equations translate directly to recent theoretical studies of synchronization in power grids''~\cite{Martens_PNAS_2013}, implying that chimera states might occur in power grids as well. 
Panaggio and Abrams also suggest that knowing the basins of stability of chimera states in power grids could be useful in avoiding them and thus maintaining the synchronous oscillation that the grid needs to function~\cite{Panaggio_Nonlinearity_2015}.
Several other chimera papers briefly refer to this possible connection~\cite{Zakharova_PRL_2014,Bohm_PRE_2015,GarciaMorales_EPL_2016,Banerjee_PRE_2016,Ulonska_Chaos_2016,Bera_EPL_2017,Rakshit_SciRep_2017}, but they seldom elaborate on it.
In fact, the article on power-grid modeling that is probably most often cited in chimera introductions, published by Motter et al. in 2013~\cite{Motter_NatPhys_2013}, contains just a single superficial reference to chimeras in its own introduction.
Its main aim is to demonstrate a condition for when network synchrony is stable and it is little concerned with what kinds of unsynchronized states may exist.
A few results are contributed by a 2014 paper by Pecora et al.~\cite{Pecora_NatComm_2014}, wherein they explain the onset of so-called isolated desynchronization by means of network topology and use two real power grids as models (among many others).
However, since isolated desynchronization is caused by topology, it should only be relevant to some kinds of chimera states (in the widest sense) and less to those arising by spontaneous symmetry breaking.

Chimera states have also been linked~\cite{Tinsley_NatPhys_2012,Sethia_PRE_2013,Schmidt_Chaos_2014,Xie_PRE_2014,Haugland_SciRep_2015,Bick_NJP_2015} to the turbulent-laminar patterns that may be observed in Taylor-Couette flow~\cite{Barkley_PRL_2005}.
Another article uses a social-agent model to suggest that ``an analogue to a chimera state'' 
could also exist in the behavior of interacting human populations~\cite{Gonzalez-Avella_PhysicaA_2014}.

At the end of a 2018 review article, 
Omel'chenko 
refers to all past attempts to identify a non-laboratory chimera as ``rather speculative'' and ``requir[ing] more rigorous justification''~\cite{OmelChenko_Nonlinearity_2018}.
With regards to most of the above examples, this indeed seems to be the case.
As far as unihemispheric sleep is concerned, we could alternatively ask exactly what kind of justification is missing.
Do actual brain-measurement data not show the coexistence of synchronized and desynchronized oscillation? 
Are these data not backed up by models modeled on the natural-world phenomenon, already declared to exhibit chimera states?
Of course, the mechanism at work in the 
bird or dolphin brain is not the same as that in all reported chimera states, but 
the latter also differ strongly among themselves.
Could we thus be holding potential natural-world chimeras to a different standard than theoretical and experimental ones?
And could this question possibly be better addressed, if the currently rather fluid and to a large extent implicit chimera definition were made more concrete?

\section{Concluding remarks}
\noindent Above, we have seen that chimera states were originally discovered in globally coupled logistic maps, earlier than often believed.
Not until they had also been produced in nonlocally coupled oscillators, however, were they discerned as a special kind of state and given a name.
Once they had a name, and once, a few years later, the versatile two-groups framework and the reference to unihemispheric sleep as a potential field of real-world relevance were introduced, a decade of expansive chimera research began.
Their number and variety increased, as did the number of systems found to support them.
Chimera states were found in a wide range of experimental settings as well.
Various derived concepts emerged, though many of them remained mostly limited to their original context.
Additional analogies to different real-world phenomena were also drawn up, though many of these have so far remained rather superficial.
Notably, the research community has not yet arrived at a common conclusion that any natural-world phenomena actually are chimera states.

This might have something to do with the fact that chimeras are not a most definite physical phenomenon (like neutrons, the Hall effect or protein folding), with definite properties we just have to measure accurately enough to discover.
The term is more abstract and might thus also be considered as a loose collection of more or less related observations.
New observations are then being given the same label as existing ones on rather discretionary grounds.
In light of all the above, this development might already have reached a point where all it requires for a type of dynamics to be called a chimera, 
is a certain minimum symmetry of the underlying system and a way of viewing it that makes one part of the depiction appear more synchronized and/or clustered than another.
This is probably both a blessing and a curse of the field, with unconstrained analogies enabling many a fruitful discovery, but at the same time counteracting the internal ordering of the entirety of results.
In particular, as ever new results have broadened the 
scope of what is called a
chimera state, there seems to have been rather limited reflection on how this has changed the object of the field itself.
Future research could probably benefit from a more explicit consideration of this insight.

\section*{Acknowledgments}
\noindent The author thanks Katharina Krischer, Felix P. Kemeth and Maximilian Patzauer for fruitful discussions.
Financial support from the Studienstiftung des deutschen Volkes 
is gratefully acknowledged.

\section*{Data availability}
\noindent Data sharing is not applicable to this article as no new data were created or analyzed in this study.


\section*{References}
\begin{thebibliography}{100}

\bibitem{Kuramoto_NPCS_2002}
Yoshiki Kuramoto and Dorjsuren Battogtokh.
\newblock {Coexistence of Coherence and Incoherence in Nonlocally Coupled Phase
  Oscillators}.
\newblock {\em Nonlinear Phenom. Complex Syst.}, 5(4):380--385, 2002.

\bibitem{Abrams_PRL_2004}
Daniel~M. Abrams and Steven~H. Strogatz.
\newblock {Chimera states for coupled oscillators}.
\newblock {\em Phys. Rev. Lett.}, 93(17):174102, 2004.

\bibitem{Sethia_PRE_2013}
Gautam~C. Sethia, Abhijit Sen, and George~L. Johnston.
\newblock {Amplitude-mediated chimera states}.
\newblock {\em Phys. Rev. E}, 88(4):042917, 2013.

\bibitem{Martens_PNAS_2013}
Erik~Andreas Martens, Shashi Thutupalli, Antoine Fourri{\`{e}}re, and Oskar
  Hallatschek.
\newblock {Chimera states in mechanical oscillator networks}.
\newblock {\em Proc. Natl. Acad. Sci. U. S. A.}, 110(26):10563--10567, 2013.

\bibitem{Nkomo_PRL_2013}
Simbarashe Nkomo, Mark~R. Tinsley, and Kenneth Showalter.
\newblock {Chimera states in populations of nonlocally coupled chemical
  oscillators}.
\newblock {\em Phys. Rev. Lett.}, 110(24):244102, 2013.

\bibitem{Omelchenko_Chaos_2015}
Iryna Omelchenko, Anna Zakharova, Philipp H{\"{o}}vel, Julien Siebert, and
  Eckehard Sch{\"{o}}ll.
\newblock {Nonlinearity of local dynamics promotes multi-chimeras}.
\newblock {\em Chaos}, 25(8):083104, 2015.

\bibitem{Ulonska_Chaos_2016}
Stefan Ulonska, Iryna Omelchenko, Anna Zakharova, and Eckehard Sch{\"{o}}ll.
\newblock {Chimera states in networks of Van der Pol oscillators with
  hierarchical connectivities}.
\newblock {\em Chaos}, 26(9):094825, 2016.

\bibitem{Schmidt_PRE_2017}
Alexander Schmidt, Theodoros Kasimatis, Johanne Hizanidis, Astero Provata, and
  Philipp H{\"{o}}vel.
\newblock {Chimera patterns in two-dimensional networks of coupled neurons}.
\newblock {\em Phys. Rev. E}, 95(3):032224, 2017.

\bibitem{Maistrenko_Chaos_2020}
Volodymyr Maistrenko, Oleksandr Sudakov, and Oleksiy Osiv.
\newblock {Chimeras and solitary states in 3D oscillator networks with
  inertia}.
\newblock {\em Chaos}, 30(6):063113, 2020.

\bibitem{Hagerstrom_NatPhys_2012}
Aaron~M. Hagerstrom, Thomas~E. Murphy, Rajarshi Roy, Philipp H{\"{o}}vel, Iryna
  Omelchenko, and Eckehard Sch{\"{o}}ll.
\newblock {Experimental observation of chimeras in coupled-map lattices}.
\newblock {\em Nat. Phys.}, 8(9):658--661, 2012.

\bibitem{Omelchenko_PRL_2011}
Iryna Omelchenko, Yuri~L. Maistrenko, Philipp H{\"{o}}vel, and Eckehard
  Sch{\"{o}}ll.
\newblock {Loss of coherence in dynamical networks: Spatial chaos and chimera
  states}.
\newblock {\em Phys. Rev. Lett.}, 106(23):234102, 2011.

\bibitem{Zakharova_PRL_2014}
Anna Zakharova, Marie Kapeller, and Eckehard Sch{\"{o}}ll.
\newblock {Chimera death: Symmetry breaking in dynamical networks}.
\newblock {\em Phys. Rev. Lett.}, 112(15):154101, 2014.

\bibitem{Zakharova_JPhysCS_2016}
Anna Zakharova, Marie Kapeller, and Eckehard Sch{\"{o}}ll.
\newblock {Amplitude chimeras and chimera death in dynamical networks}.
\newblock {\em J. Phys. Conf. Ser.}, 727(1):012018, 2016.

\bibitem{Schmidt_Chaos_2014}
Lennart Schmidt, Konrad Sch{\"{o}}nleber, Katharina Krischer, and Vladimir
  Garc{\'{i}}a-Morales.
\newblock {Coexistence of synchrony and incoherence in oscillatory media under
  nonlinear global coupling}.
\newblock {\em Chaos}, 24(1):013102, 2014.

\bibitem{Schmidt_Chaos_2015}
Lennart Schmidt and Katharina Krischer.
\newblock {Chimeras in globally coupled oscillatory systems: From ensembles of
  oscillators to spatially continuous media}.
\newblock {\em Chaos}, 25(6):64401, jun 2015.

\bibitem{Bohm_PRE_2015}
Fabian B{\"{o}}hm, Anna Zakharova, Eckehard Sch{\"{o}}ll, and Kathy
  L{\"{u}}dge.
\newblock {Amplitude-phase coupling drives chimera states in globally coupled
  laser networks}.
\newblock {\em Phys. Rev. E}, 91(4):040901, 2015.

\bibitem{Rohm_PRE_2016}
Andr{\'{e}} R{\"{o}}hm, Fabian B{\"{o}}hm, and Kathy L{\"{u}}dge.
\newblock {Small chimera states without multistability in a globally
  delay-coupled network of four lasers}.
\newblock {\em Phys. Rev. E}, 94(4):042204, 2016.

\bibitem{Cho_PRL_2017}
Young~Sul Cho, Takashi Nishikawa, and Adilson~E. Motter.
\newblock {Stable Chimeras and Independently Synchronizable Clusters}.
\newblock {\em Phys. Rev. Lett.}, 119(8):084101, 2017.

\bibitem{OmelChenko_Nonlinearity_2018}
Oleh~E. Omel'chenko.
\newblock {The mathematics behind chimera states}.
\newblock {\em Nonlinearity}, 31(5):R121--R164, 2018.

\bibitem{Kaneko_Chaos_2015}
Kunihiko Kaneko.
\newblock {From globally coupled maps to complex-systems biology}.
\newblock {\em Chaos}, 25(9):097608, 2015.

\bibitem{Kemeth_Chaos_2016}
Felix~P. Kemeth, Sindre~W. Haugland, Lennart Schmidt, Ioannis~G. Kevrekidis,
  and Katharina Krischer.
\newblock {A classification scheme for chimera states}.
\newblock {\em Chaos}, 26(9):094815, 2016.

\bibitem{Hart_Chaos_2016}
Joseph~D. Hart, Kanika Bansal, Thomas~E. Murphy, and Rajarshi Roy.
\newblock {Experimental observation of chimera and cluster states in a minimal
  globally coupled network}.
\newblock {\em Chaos}, 26(9):094801, 2016.

\bibitem{Kaneko_PhysicaD_1990}
Kunihiko Kaneko.
\newblock {Clustering, coding, switching, hierarchical ordering, and control in
  a network of chaotic elements}.
\newblock {\em Phys. D Nonlinear Phenom.}, 41(2):137--172, 1990.

\bibitem{Golomb_PRA_1992}
D.~Golomb, D.~Hansel, B.~Shraiman, and H.~Sompolinsky.
\newblock {Clustering in globally coupled phase oscillators}.
\newblock {\em Phys. Rev. A}, 45(6):3516--3530, 1992.

\bibitem{Okuda_PhysicaD_1993}
Koji Okuda.
\newblock {Variety and generality of clustering in globally coupled
  oscillators}.
\newblock {\em Phys. D Nonlinear Phenom.}, 63(3-4):424--436, 1993.

\bibitem{Hakim_PRA_1992}
Vincent Hakim and Wouter~Jan Rappel.
\newblock {Dynamics of the globally coupled complex Ginzburg-Landau equation}.
\newblock {\em Phys. Rev. A}, 46(12):R7347--R7350, 1992.

\bibitem{Nakagawa_PTPS_1993}
Naoko Nakagawa and Yoshiki Kuramoto.
\newblock {Collective Chaos in a Population of Globally Coupled Oscillators}.
\newblock {\em Prog. Theor. Phys.}, 89(2):313--323, 1993.

\bibitem{Panaggio_Nonlinearity_2015}
Mark~J. Panaggio and Daniel~M. Abrams.
\newblock {Chimera states: Coexistence of coherence and incoherence in networks
  of coupled oscillators}.
\newblock {\em Nonlinearity}, 28(3):R67--R87, 2015.

\bibitem{Scholl_EPJST_2016}
Eckehard Sch{\"{o}}ll.
\newblock {Synchronization patterns and chimera states in complex networks:
  Interplay of topology and dynamics}.
\newblock {\em Eur. Phys. J. Spec. Top.}, 225(6-7):891--919, 2016.

\bibitem{Bera_EPL_2017}
Bidesh~K. Bera, Soumen Majhi, Dibakar Ghosh, and Matja{\v{z}} Perc.
\newblock {Chimera states: Effects of different coupling topologies}.
\newblock {\em Europhys. Lett.}, 118(1):10001, 2017.

\bibitem{Sethia_PRL_2014}
Gautam~C. Sethia and Abhijit Sen.
\newblock {Chimera states: The existence criteria revisited}.
\newblock {\em Phys. Rev. Lett.}, 112(14):144101, 2014.

\bibitem{Ma_EPL_2010}
Rubao Ma, Jianxiong Wang, and Zonghua Liu.
\newblock {Robust features of chimera states and the implementation of
  alternating chimera states}.
\newblock {\em Europhys. Lett.}, 91(4):40006, 2010.

\bibitem{Omelchenko_PRL_2013}
Iryna Omelchenko, Oleh~E. Omel'Chenko, Philipp H{\"{o}}vel, and Eckehard
  Sch{\"{o}}ll.
\newblock {When nonlocal coupling between oscillators becomes stronger: Patched
  synchrony or multichimera states}.
\newblock {\em Phys. Rev. Lett.}, 110(22):224101, 2013.

\bibitem{Ashwin_Chaos_2015}
Peter Ashwin and Oleksandr Burylko.
\newblock {Weak chimeras in minimal networks of coupled phase oscillators}.
\newblock {\em Chaos}, 25(1):013106, 2015.

\bibitem{Shima_PRE_2004}
Shin~Ichiro Shima and Yoshiki Kuramoto.
\newblock {Rotating spiral waves with phase-randomized core in nonlocally
  coupled oscillators}.
\newblock {\em Phys. Rev. E}, 69(3 2):036213, 2004.

\bibitem{Abrams_IJBC_2006}
Daniel~M. Abrams and Steven~H. Strogatz.
\newblock {Chimera States in a Ring of Nonlocally Coupled Oscillators Daniel}.
\newblock {\em Int. J. Bifurc. Chaos}, 16(01):21--37, 2006.

\bibitem{Sethia_PRL_2008}
Gautam~C. Sethia, Abhijit Sen, and Fatihcan~M. Atay.
\newblock {Clustered chimera states in delay-coupled oscillator systems}.
\newblock {\em Phys. Rev. Lett.}, 100(14):144102, 2008.

\bibitem{Wolfrum_Chaos_2011}
Matthias Wolfrum, Oleh~E. Omel'Chenko, Serhiy Yanchuk, and Yuri~L. Maistrenko.
\newblock {Spectral properties of chimera states}.
\newblock {\em Chaos}, 21(1):013112, 2011.

\bibitem{Wolfrum_PRE_2011}
Matthias Wolfrum and Oleh~E. Omel'Chenko.
\newblock {Chimera states are chaotic transients}.
\newblock {\em Phys. Rev. E}, 84(1):015201, 2011.

\bibitem{Kawamura_PRE_2007}
Yoji Kawamura.
\newblock {Chimera Ising walls in forced nonlocally coupled oscillators}.
\newblock {\em Phys. Rev. E}, 75(5):056204, 2007.

\bibitem{OmelChenko_PRL_2008}
Oleh~E. Omel'chenko, Yuri~L. Maistrenko, and Peter~A. Tass.
\newblock {Chimera states: The natural link between coherence and incoherence}.
\newblock {\em Phys. Rev. Lett.}, 100(4):044105, 2008.

\bibitem{Abrams_PRL_2008}
Daniel~M. Abrams, Rennie Mirollo, Steven~H. Strogatz, and Daniel~A. Wiley.
\newblock {Solvable model for chimera states of coupled oscillators}.
\newblock {\em Phys. Rev. Lett.}, 101(8):84103, 2008.

\bibitem{Ott_Chaos_2008}
Edward Ott and Thomas~M. Antonsen.
\newblock {Low dimensional behavior of large systems of globally coupled
  oscillators}.
\newblock {\em Chaos}, 18(3):037113, 2008.

\bibitem{Pikovsky_PRL_2008}
Arkady Pikovsky and Michael Rosenblum.
\newblock {Partially Integrable Dynamics of Hierarchical Populations of Coupled
  Oscillators}.
\newblock {\em Phys. Rev. Lett.}, 101(26):264103, 2008.

\bibitem{Laing_Chaos_2009}
Carlo~R. Laing.
\newblock {Chimera states in heterogeneous networks}.
\newblock {\em Chaos An Interdiscip. J. Nonlinear Sci.}, 19(1):013113, 2009.

\bibitem{Laing_PhysicaD_2009}
Carlo~R. Laing.
\newblock {The dynamics of chimera states in heterogeneous Kuramoto networks}.
\newblock {\em Phys. D Nonlinear Phenom.}, 238(16):1569--1588, 2009.

\bibitem{Sheeba_PRE_2009}
Jane~H. Sheeba, V.~K. Chandrasekar, and M.~Lakshmanan.
\newblock {Globally clustered chimera states in delay-coupled populations}.
\newblock {\em Phys. Rev. E}, 79(5):055203, 2009.

\bibitem{Sheeba_PRE_2010}
Jane~H. Sheeba, V.~K. Chandrasekar, and M.~Lakshmanan.
\newblock {Chimera and globally clustered chimera: Impact of time delay}.
\newblock {\em Phys. Rev. E}, 81(4):046203, 2010.

\bibitem{Bordyugov_PRE_2010}
Grigory Bordyugov, Arkady Pikovsky, and Michael Rosenblum.
\newblock {Self-emerging and turbulent chimeras in oscillator chains}.
\newblock {\em Phys. Rev. E}, 82(3):035205, 2010.

\bibitem{Martens_PRL_2010}
Erik~A. Martens, Carlo~R. Laing, and Steven~H. Strogatz.
\newblock {Solvable model of spiral wave chimeras}.
\newblock {\em Phys. Rev. Lett.}, 104(4):044101, 2010.

\bibitem{Martens_PRE_2010}
Erik~A. Martens.
\newblock {Bistable chimera attractors on a triangular network of oscillator
  populations}.
\newblock {\em Phys. Rev. E}, 82(1):016216, 2010.

\bibitem{Martens_Chaos_2010}
Erik~A. Martens.
\newblock {Chimeras in a network of three oscillator populations with varying
  network topology}.
\newblock {\em Chaos}, 20(4):043122, 2010.

\bibitem{Shanahan_Chaos_2010}
Murray Shanahan.
\newblock {Metastable chimera states in community-structured oscillator
  networks}.
\newblock {\em Chaos}, 20(1):013108, 2010.

\bibitem{OmelChenko_PRE_2010}
Oleh~E. Omel'chenko, Matthias Wolfrum, and Yuri~L. Maistrenko.
\newblock {Chimera states as chaotic spatiotemporal patterns}.
\newblock {\em Phys. Rev. E}, 81(6):065201, 2010.

\bibitem{Laing_Chaos_2012}
Carlo~R. Laing, Karthikeyan Rajendran, and Ioannis~G. Kevrekidis.
\newblock {Chimeras in random non-complete networks of phase oscillators}.
\newblock {\em Chaos}, 22(1):013132, 2012.

\bibitem{Wildie_Chaos_2012}
Mark Wildie and Murray Shanahan.
\newblock {Metastability and chimera states in modular delay and pulse-coupled
  oscillator networks}.
\newblock {\em Chaos}, 22(4):043131, 2012.

\bibitem{Laing_PRE_2010}
Carlo~R. Laing.
\newblock {Chimeras in networks of planar oscillators}.
\newblock {\em Phys. Rev. E}, 81(6):066221, 2010.

\bibitem{Laing_PRE_2015}
Carlo~R. Laing.
\newblock {Chimeras in networks with purely local coupling}.
\newblock {\em Phys. Rev. E}, 92(5):050904, 2015.

\bibitem{Laing_Chaos_2012a}
Carlo~R. Laing.
\newblock {Disorder-induced dynamics in a pair of coupled heterogeneous phase
  oscillator networks}.
\newblock {\em Chaos}, 22(4):043104, 2012.

\bibitem{Tinsley_NatPhys_2012}
Mark~R. Tinsley, Simbarashe Nkomo, and Kenneth Showalter.
\newblock {Chimera and phase-cluster states in populations of coupled chemical
  oscillators}.
\newblock {\em Nat. Phys.}, 8(9):662--665, 2012.

\bibitem{Pazo_PRX_2014}
Diego Paz{\'{o}} and Ernest Montbri{\'{o}}.
\newblock {Low-Dimensional Dynamics of Populations of Pulse-Coupled
  Oscillators}.
\newblock {\em Phys. Rev. X}, 4(1):011009, 2014.

\bibitem{Buscarino_PRE_2015}
Arturo Buscarino, Mattia Frasca, Lucia~Valentina Gambuzza, and Philipp
  H{\"{o}}vel.
\newblock {Chimera states in time-varying complex networks}.
\newblock {\em Phys. Rev. E}, 91(2):022817, 2015.

\bibitem{Panaggio_PRE_2016}
Mark~J. Panaggio, Daniel~M. Abrams, Peter Ashwin, and Carlo~R. Laing.
\newblock {Chimera states in networks of phase oscillators: The case of two
  small populations}.
\newblock {\em Phys. Rev. E}, 93(1):012218, 2016.

\bibitem{Martens_Chaos_2016}
Erik~A. Martens, Christian Bick, and Mark~J. Panaggio.
\newblock {Chimera states in two populations with heterogeneous phase-lag}.
\newblock {\em Chaos}, 26(9):094819, 2016.

\bibitem{Martens_NJP_2016}
Erik~A. Martens, Mark~J. Panaggio, and Daniel~M. Abrams.
\newblock {Basins of attraction for chimera states}.
\newblock {\em New J. Phys.}, 18(2):022002, 2016.

\bibitem{Montbrio_PRE_2004}
Ernest Montbri{\'{o}}, J{\"{u}}rgen Kurths, and Bernd Blasius.
\newblock {Synchronization of two interacting populations of oscillators}.
\newblock {\em Phys. Rev. E}, 70(5):056125, 2004.

\bibitem{Barreto_PRE_2008}
Ernest Barreto, Brian Hunt, Edward Ott, and Paul So.
\newblock {Synchronization in networks of networks: The onset of coherent
  collective behavior in systems of interacting populations of heterogeneous
  oscillators}.
\newblock {\em Phys. Rev. E}, 77(3):036107, 2008.

\bibitem{Vuellings_NJP_2014}
Andrea V{\"{u}}llings, Johanne Hizanidis, Iryna Omelchenko, and Philipp
  H{\"{o}}vel.
\newblock {Clustered chimera states in systems of type-I excitability}.
\newblock {\em New J. Phys.}, 16(12):123039, 2014.

\bibitem{Ujjwal_PRE_2013}
Sangeeta~Rani Ujjwal and Ramakrishna Ramaswamy.
\newblock {Chimeras with multiple coherent regions}.
\newblock {\em Phys. Rev. E}, 88(3):032902, 2013.

\bibitem{Xie_PRE_2014}
Jianbo Xie, Edgar Knobloch, and Hsien-Ching Kao.
\newblock {Multicluster and traveling chimera states in nonlocal phase-coupled
  oscillators}.
\newblock {\em Phys. Rev. E}, 90(2):022919, 2014.

\bibitem{Yao_SciRep_2015}
Nan Yao, Zi~Gang Huang, Celso Grebogi, and Ying~Cheng Lai.
\newblock {Emergence of multicluster chimera states}.
\newblock {\em Sci. Rep.}, 5:12988, 2015.

\bibitem{Maistrenko_IJBC_2014}
Yuri~L. Maistrenko, Anna Vasylenko, Oleksandr Sudakov, Roman Levchenko, and
  Volodymyr~L. Maistrenko.
\newblock {Cascades of multiheaded chimera states for coupled phase
  oscillators}.
\newblock {\em Int. J. Bifurc. Chaos}, 24(8):1440014, 2014.

\bibitem{Larger_NatCom_2015}
Laurent Larger, Bogdan Penkovsky, and Yuri Maistrenko.
\newblock {Laser chimeras as a paradigm for multistable patterns in complex
  systems}.
\newblock {\em Nat. Commun.}, 6:7752, 2015.

\bibitem{Hizanidis_PRE_2015}
Johanne Hizanidis, Evangelia Panagakou, Iryna Omelchenko, Eckehard
  Sch{\"{o}}ll, Philipp H{\"{o}}vel, and Astero Provata.
\newblock {Chimera states in population dynamics: Networks with fragmented and
  hierarchical connectivities}.
\newblock {\em Phys. Rev. E}, 92(1):012915, 2015.

\bibitem{Xie_PRE_2015}
Jianbo Xie, Edgar Knobloch, and Hsien-Ching Kao.
\newblock {Twisted chimera states and multicore spiral chimera states on a
  two-dimensional torus}.
\newblock {\em Phys. Rev. E}, 92(4):042921, 2015.

\bibitem{Haugland_SciRep_2015}
Sindre~W. Haugland, Lennart Schmidt, and Katharina Krischer.
\newblock {Self-organized alternating chimera states in oscillatory media}.
\newblock {\em Sci. Rep.}, 5:9883, 2015.

\bibitem{Suda_PRE_2015}
Yusuke Suda and Koji Okuda.
\newblock {Persistent chimera states in nonlocally coupled phase oscillators}.
\newblock {\em Phys. Rev. E}, 92(6):060901, 2015.

\bibitem{Omelchenko_PRL_2016}
Iryna Omelchenko, Oleh~E. Omel'chenko, Anna Zakharova, Matthias Wolfrum, and
  Eckehard Sch{\"{o}}ll.
\newblock {Tweezers for Chimeras in Small Networks}.
\newblock {\em Phys. Rev. Lett.}, 116(11):114101, 2016.

\bibitem{GarciaMorales_EPL_2016}
Vladimir Garc{\'{i}}a-Morales.
\newblock {Cellular automaton for chimera states}.
\newblock {\em Europhys. Lett.}, 114(1):18002, 2016.

\bibitem{Maistrenko_PRE_2017}
Yuri Maistrenko, Serhiy Brezetsky, Patrycja Jaros, Roman Levchenko, and Tomasz
  Kapitaniak.
\newblock {Smallest chimera states}.
\newblock {\em Phys. Rev. E}, 95(1):010203, 2017.

\bibitem{Kemeth_PRL_2018}
Felix~P. Kemeth, Sindre~W. Haugland, and Katharina Krischer.
\newblock {Symmetries of Chimera States}.
\newblock {\em Phys. Rev. Lett.}, 120(21):214101, 2018.

\bibitem{OcampoEspindola_FAMS_2019}
Jorge~Luis Ocampo-Espindola, Christian Bick, and Istv{\'{a}}n~Z. Kiss.
\newblock {Weak Chimeras in Modular Electrochemical Oscillator Networks}.
\newblock {\em Front. Appl. Math. Stat.}, 5:38, 2019.

\bibitem{Rosin_PRE_2014}
David~P. Rosin, Damien Rontani, Nicholas~D. Haynes, Eckehard Sch{\"{o}}ll, and
  Daniel~J. Gauthier.
\newblock {Transient scaling and resurgence of chimera states in networks of
  Boolean phase oscillators}.
\newblock {\em Phys. Rev. E}, 90(3):030902, 2014.

\bibitem{Qiu_SciRep_2016}
Tian Qiu, Stefano Boccaletti, Ivan Bonamassa, Yong Zou, Jie Zhou, Zonghua Liu,
  and Shuguang Guan.
\newblock {Synchronization and Bellerophon states in conformist and contrarian
  oscillators}.
\newblock {\em Sci. Rep.}, 6(1):36713, 2016.

\bibitem{Maistrenko_PRE_2014}
Yuri Maistrenko, Bogdan Penkovsky, and Michael Rosenblum.
\newblock {Solitary state at the edge of synchrony in ensembles with attractive
  and repulsive interactions}.
\newblock {\em Phys. Rev. E}, 89(6):060901, 2014.

\bibitem{Jaros_PRE_2015}
Patrycja Jaros, Yuri Maistrenko, and Tomasz Kapitaniak.
\newblock {Chimera states on the route from coherence to rotating waves}.
\newblock {\em Phys. Rev. E}, 91(2):1--5, 2015.

\bibitem{Olmi_PRE_2015}
Simona Olmi, Erik~A. Martens, Shashi Thutupalli, and Alessandro Torcini.
\newblock {Intermittent chaotic chimeras for coupled rotators}.
\newblock {\em Phys. Rev. E}, 92(3):030901, 2015.

\bibitem{Maistrenko_NJP_2015}
Yuri Maistrenko, Oleksandr Sudakov, Oleksiy Osiv, and Volodymyr Maistrenko.
\newblock {Chimera states in three dimensions}.
\newblock {\em New J. Phys.}, 17(7):073037, 2015.

\bibitem{Goldschmidt_Chaos_2019}
Richard~Janis Goldschmidt, Arkady Pikovsky, and Antonio Politi.
\newblock {Blinking chimeras in globally coupled rotators}.
\newblock {\em Chaos}, 29(7):071101, 2019.

\bibitem{Petrungaro_PRE_2017}
Gabriela Petrungaro, Koichiro Uriu, and Luis~G. Morelli.
\newblock {Mobility-induced persistent chimera states}.
\newblock {\em Phys. Rev. E}, 96(6):062210, 2017.

\bibitem{Battogtokh_PhysicaD_1997}
D.~Battogtokh, A.~Preusser, and A.~Mikhailov.
\newblock {Controlling turbulence in the complex Ginzburg-Landau equation II.
  Two-dimensional systems}.
\newblock {\em Phys. D Nonlinear Phenom.}, 106(3-4):327--362, 1997.

\bibitem{Falcke_PRE_1994}
Martin Falcke and Harald Engel.
\newblock {Influence of global coupling through the gas phase on the dynamics
  of CO oxidation on Pt(110)}.
\newblock {\em Phys. Rev. E}, 50(2):1353--1359, 1994.

\bibitem{Falcke_JChemPhys_1994}
Martin Falcke and Harald Engel.
\newblock {Pattern formation during the CO oxidation on Pt(110) surfaces under
  global coupling}.
\newblock {\em J. Chem. Phys.}, 101(7):6255--6263, 1994.

\bibitem{Falcke_Thesis_1995}
Martin Falcke.
\newblock {\em {Strukturbildung in Reaktions- Diffusionssystemen und globale
  Kopplung}}.
\newblock Wiss.-und-Technik-Verlag Gross, Berlin, 1995.

\bibitem{Yang_PRE_2000}
Lingfa Yang, Milos Dolnik, Anatol~M. Zhabotinsky, and Irving~R. Epstein.
\newblock {Oscillatory clusters in a model of the photosensitive
  Belousov-Zhabotinsky reaction system with global feedback}.
\newblock {\em Phys. Rev. E}, 62(5):6414--6420, 2000.

\bibitem{Note1}
While their 2012 paper simply refers to ``the projected image from a spatial
  light modulator''~\cite {Tinsley_NatPhys_2012}, a later article by the same
  authors on a very similar experiment mentions that ``[t]he experimental
  set-up consists of a modified video projector (SLM) with a 440–460 nm band
  pass filter''~\cite {Nkomo_Chaos_2016}. Thus it is not unlikely that the SLM
  used in the 2012 experiment was this custom-built apparatus as well.

\bibitem{Smart_PhysToday_2012}
Ashley~G. Smart.
\newblock {Exotic chimera dynamics glimpsed in experiments}.
\newblock {\em Phys. Today}, 65(10):17--19, 2012.

\bibitem{Wojewoda_SciRep_2016}
Jerzy Wojewoda, Krzysztof Czolczynski, Yuri Maistrenko, and Tomasz Kapitaniak.
\newblock {The smallest chimera state for coupled pendula}.
\newblock {\em Sci. Rep.}, 6(1):34329, 2016.

\bibitem{Wickramasinghe_PlosOne_2013}
Mahesh Wickramasinghe and Istv{\'{a}}n~Z. Kiss.
\newblock {Spatially organized dynamical states in chemical oscillator
  networks: Synchronization, dynamical differentiation, and chimera patterns}.
\newblock {\em PLoS One}, 8(11):e80586, 2013.

\bibitem{Gambuzza_PRE_2014}
Lucia~Valentina Gambuzza, Arturo Buscarino, Sergio Chessari, Luigi Fortuna,
  Riccardo Meucci, and Mattia Frasca.
\newblock {Experimental investigation of chimera states with quiescent and
  synchronous domains in coupled electronic oscillators}.
\newblock {\em Phys. Rev. E}, 90(3):032905, 2014.

\bibitem{Brunner_Chaos_2018}
D.~Brunner, B.~Penkovsky, R.~Levchenko, E.~Sch{\"{o}}ll, L.~Larger, and
  Y.~Maistrenko.
\newblock {Two-dimensional spatiotemporal complexity in dual-delayed nonlinear
  feedback systems: Chimeras and dissipative solitons}.
\newblock {\em Chaos}, 28(10):103106, 2018.

\bibitem{Schoenleber_NJP_2014}
Konrad Sch{\"{o}}nleber, Carla Zensen, Andreas Heinrich, and Katharina
  Krischer.
\newblock {Pattern formation during the oscillatory photoelectrodissolution of
  n-type silicon: Turbulence, clusters and chimeras}.
\newblock {\em New J. Phys.}, 16(6):063024, 2014.

\bibitem{Totz_NatPhys_2017}
Jan~Frederik Totz, Julian Rode, Mark~R. Tinsley, Kenneth Showalter, and Harald
  Engel.
\newblock {Spiral wave chimera states in large populations of coupled chemical
  oscillators}.
\newblock {\em Nat. Phys.}, 14(3):282--285, 2018.

\bibitem{Bastidas_PRE_2015}
V.~M. Bastidas, I.~Omelchenko, A.~Zakharova, E.~Sch{\"{o}}ll, and T.~Brandes.
\newblock {Quantum signatures of chimera states}.
\newblock {\em Phys. Rev. E}, 92(6):062924, 2015.

\bibitem{Bastidas_Book_2016}
Victor~Manuel Bastidas, Iryna Omelchenko, Anna Zakharova, Eckehard
  Sch{\"{o}}ll, and Tobias Brandes.
\newblock {Chimera States in Quantum Mechanics}.
\newblock In Eckehard Sch{\"{o}}ll, Sabine H.~L. Klapp, and Philipp
  H{\"{o}}vel, editors, {\em Control of Self-Organizing Nonlinear Systems}.
  Springer, Cham, 2016.

\bibitem{Mukhametov_BrainRes_1977}
L.~M. Mukhametov, A.~Y. Supin, and I.~G. Polyakova.
\newblock {Interhemispheric asymmetry of the electroencephalographic sleep
  patterns in dolphins}.
\newblock {\em Brain Res.}, 134(3):581--584, 1977.

\bibitem{Rattenborg_NeuBioRev_2000}
N.~C. Rattenborg, C.~J. Amlaner, and S.~L. Lima.
\newblock {Behavioral, neurophysiological and evolutionary perspectives on
  unihemispheric sleep}.
\newblock {\em Neurosci. Biobehav. Rev.}, 24(8):817--842, 2000.

\bibitem{Mathews_Ethology_2006}
Christian~G. Mathews, John~A. Lesku, Steven~L. Lima, and Charles~J. Amlaner.
\newblock {Asynchronous eye closure as an anti-predator behavior in the western
  fence lizard (Sceloporus occidentalis)}.
\newblock {\em Ethology}, 112(3):286--292, 2006.

\bibitem{Ramlow_EPL_2019}
Lukas Ramlow, Jakub Sawicki, Anna Zakharova, Jaroslav Hlinka, Jens~Christian
  Claussen, and Eckehard Sch{\"{o}}ll.
\newblock {Partial synchronization in empirical brain networks as a model for
  unihemispheric sleep}.
\newblock {\em Europhys. Lett.}, 126(5):50007, 2019.

\bibitem{Kang_SciRep_2019}
Ling Kang, Changhai Tian, Siyu Huo, and Zonghua Liu.
\newblock {A two-layered brain network model and its chimera state}.
\newblock {\em Sci. Rep.}, 9(1):14389, 2019.

\bibitem{Tamaki_CurrBio_2016}
Masako Tamaki, Ji~Won Bang, Takeo Watanabe, and Yuka Sasaki.
\newblock {Night Watch in One Brain Hemisphere during Sleep Associated with the
  First-Night Effect in Humans}.
\newblock {\em Curr. Biol.}, 26(9):1190--1194, 2016.

\bibitem{Banerjee_PRE_2016}
Tanmoy Banerjee, Partha~Sharathi Dutta, Anna Zakharova, and Eckehard
  Sch{\"{o}}ll.
\newblock {Chimera patterns induced by distance-dependent power-law coupling in
  ecological networks}.
\newblock {\em Phys. Rev. E}, 94(3):032206, 2016.

\bibitem{Rakshit_SciRep_2017}
Sarbendu Rakshit, Bidesh~K. Bera, Matja{\v{z}} Perc, and Dibakar Ghosh.
\newblock {Basin stability for chimera states}.
\newblock {\em Sci. Rep.}, 7(1):2412, 2017.

\bibitem{Motter_NatPhys_2013}
Adilson~E. Motter, Seth~A. Myers, Marian Anghel, and Takashi Nishikawa.
\newblock {Spontaneous synchrony in power-grid networks}.
\newblock {\em Nat. Phys.}, 9(3):191--197, 2013.

\bibitem{Pecora_NatComm_2014}
Louis~M. Pecora, Francesco Sorrentino, Aaron~M. Hagerstrom, Thomas~E. Murphy,
  and Rajarshi Roy.
\newblock {Cluster synchronization and isolated desynchronization in complex
  networks with symmetries}.
\newblock {\em Nat. Commun.}, 5:4079, 2014.

\bibitem{Bick_NJP_2015}
Christian Bick and Erik~A. Martens.
\newblock {Controlling chimeras}.
\newblock {\em New J. Phys.}, 17(3):033030, 2015.

\bibitem{Barkley_PRL_2005}
Dwight Barkley and Laurette~S. Tuckerman.
\newblock {Computational study of turbulent laminar patterns in couette flow}.
\newblock {\em Phys. Rev. Lett.}, 94(1):14502, 2005.

\bibitem{Gonzalez-Avella_PhysicaA_2014}
J.~C. Gonz{\'{a}}lez-Avella, M.~G. Cosenza, and M.~{San Miguel}.
\newblock {Localized coherence in two interacting populations of social
  agents}.
\newblock {\em Phys. A Stat. Mech. its Appl.}, 399:24--30, 2014.

\bibitem{Nkomo_Chaos_2016}
Simbarashe Nkomo, Mark~R. Tinsley, and Kenneth Showalter.
\newblock {Chimera and chimera-like states in populations of nonlocally coupled
  homogeneous and heterogeneous chemical oscillators}.
\newblock {\em Chaos An Interdiscip. J. Nonlinear Sci.}, 26(9):094826, 2016.

\end{thebibliography}
\end{document}